\documentclass[10pt,conference]{IEEEtran}
\IEEEoverridecommandlockouts

\usepackage{hyperref}

\usepackage[table,xcdraw]{xcolor}
\usepackage{multirow}

\usepackage{cite}

\usepackage{amsmath,amssymb,amsfonts}
\usepackage{algorithmic}
\usepackage{graphicx}
\usepackage{textcomp}
\usepackage{xcolor}
\def\BibTeX{{\rm B\kern-.05em{\sc i\kern-.025em b}\kern-.08em
    T\kern-.1667em\lower.7ex\hbox{E}\kern-.125emX}}

\newcolumntype{M}[1]{>{\centering\arraybackslash}m{#1}}

\usepackage{flushend}

\usepackage{framed}
\colorlet{shadecolor}{gray!20}

\begin{document}

\title{Software Development During COVID-19 Pandemic: an Analysis of Stack Overflow and GitHub}

\author{
\IEEEauthorblockN{Pedro Almir M. Oliveira}
\IEEEauthorblockA{
  \textit{Group of Computer Networks, Software}     \\ 
  \textit{Engineering and Systems (GREat)}          \\ 
  \textit{Federal University of Cear\'{a} (UFC)}    \\
  Fortaleza, Cear\'{a}, Brazil                       \\
  pedromartins@great.ufc.br
}
\and
\IEEEauthorblockN{Pedro A. Santos Neto}
\IEEEauthorblockA{
  \textit{Laboratory of Software}                   \\
  \textit{Optimization and Testing (LOST)}          \\
  \textit{Federal University of Piau\'{i} (UFPI)}   \\ 
  Teresina, Piau\'{i}, Brazil                       \\
  pasn@ufpi.edu.br
}
\and
\IEEEauthorblockN{Gleison Silva}
\IEEEauthorblockA{
  \textit{Maida Health Company}     \\
  Teresina, Piau\'{i}, Brazil       \\
  gleison@infoway-pi.com.br         \\
  https://maida.health
}
\and
\IEEEauthorblockN{Irvayne Ibiapina}
\IEEEauthorblockA{
  \textit{Laboratory of Software}                   \\
  \textit{Optimization and Testing (LOST)}          \\
  \textit{Federal University of Piau\'{i} (UFPI)}   \\ 
  Teresina, Piau\'{i}, Brazil                       \\
  irvaynematheus@gmail.com
}
\and
\IEEEauthorblockN{Werney L. Lira}
\IEEEauthorblockA{
  \textit{Federal Institute of Education,}              \\
  \textit{Science, and Technology of Piau\'{i} (IFPI)}  \\
  Pedro II, Piau\'{i}, Brazil                           \\
  werney@ifpi.edu.br
}
\and
\IEEEauthorblockN{Rossana M. C. Andrade}
\IEEEauthorblockA{
  \textit{Group of Computer Networks, Software}     \\ 
  \textit{Engineering and Systems (GREat)}          \\ 
  \textit{Federal University of Cear\'{a} (UFC)}    \\
  Fortaleza, Cear\'{a}, Brazil                       \\
  rossana@great.ufc.br
}
}

\maketitle

\begin{abstract}
The new coronavirus became a severe health issue for the world. This situation has motivated studies of different areas to combat this pandemic. In software engineering, we point out data visualization projects to follow the disease evolution, machine learning to estimate the pandemic behavior, and computer vision processing radiologic images. Most of these projects are stored in version control systems, and there are discussions about them in Question \& Answer websites. In this work, we conducted a Mining Software Repository on a large number of questions and projects aiming to find trends that could help researchers and practitioners to fight against the coronavirus. We analyzed 1,190 questions from Stack Overflow and Data Science Q\&A and 60,352 GitHub projects. We identified a correlation between the questions and projects throughout the pandemic. The main questions about coronavirus are how-to, related to web scraping and data visualization, using Python, JavaScript, and R. The most recurrent GitHub projects are machine learning projects, using JavaScript, Python, and Java.
\end{abstract}


\section{Introduction}
\label{sec:intro}
The world is facing a new problem that has emerged in 2019. The COVID-19 pandemic is an ongoing pandemic caused by the coronavirus SARS-CoV-2 \cite{munster2020novel}. The outbreak was first identified in Wuhan, China, in December 2019. The World Health Organization (WHO) declared the outbreak a Public Health Emergency of International Concern on 30 January and a pandemic on 11 March. The first confirmed and recorded cases by WHO date from 11 January 2020. As of 12 January 2021, more than 91 million cases of COVID-19 have been reported in more than 188 countries and territories, resulting in more than 1.9 million of deaths.

The crisis caused by the coronavirus has stimulated an excellent human characteristic: helpfulness. Many people are trying to help in different ways. In particular, there are many people developing software that can somehow assist in fighting the new coronavirus. These initiatives are generating projects, which are being stored in version control systems, while numerous questions about software development are being submitted to various Question \& Answer (Q\&A) websites.

In this work, we tried to investigate at a higher level of depth what the software development community is doing to help fight the coronavirus. To the best of our knowledge, we did not find papers aiming to analyze this kind of information. Our idea then was to perform such an investigation in one of the most famous Question \& Answer website at the moment, the Stack Overflow (SO)\footnote{Stack Overflow website: \href{http://stackoverflow.com/}{http://stackoverflow.com/}.}, as well as using one of the most used software project repository today, GitHub (GH)\footnote{GitHub website: \href{https://github.com/}{https://github.com/}.}.

The following research questions guided this work:

\begin{itemize}
    \item \textbf{RQ1}: What is the temporal distribution of questions and projects considering the COVID-19 pandemic timeline?
    \item \textbf{RQ2}: What are the characteristics of the questions asked by developers in the context of the COVID-19?
    \item \textbf{RQ3}: Which programming languages have been used to develop solutions related to the COVID-19?
    \item \textbf{RQ4}: What are the hot topics considering the questions and project descriptions in the context of the COVID-19 pandemic?
    \item \textbf{RQ5}: Which GitHub projects related to the COVID-19 context stand out?
\end{itemize}

We highlight that this paper brings contributions to practitioners and researchers as it describes the main topics related to the development of COVID-19 solutions while providing information on key issues identified, facilitating access, and thus serving as a source of assistance for the development of this kind of solutions. Also, this paper present a broad overview of software community behavior throughout the beginning of COVID-19 pandemic and this knowledge can be used to guide further investigations about how to enhance this behavior or to create actions plans for future similar situations.

\section{Methodology}
\label{sec:method}
This paper reports a Mining Software Repository (MSR) \cite{hemmati2013msr} to identify trends and opportunities in the software development related to  COVID-19. Thus, we have designed our methodology following the guidelines proposed by \cite{hemmati2013msr} and \cite{kitchenham2015evidence}.

\begin{figure}[h]
    \centering
    \includegraphics[width = 0.8\columnwidth]{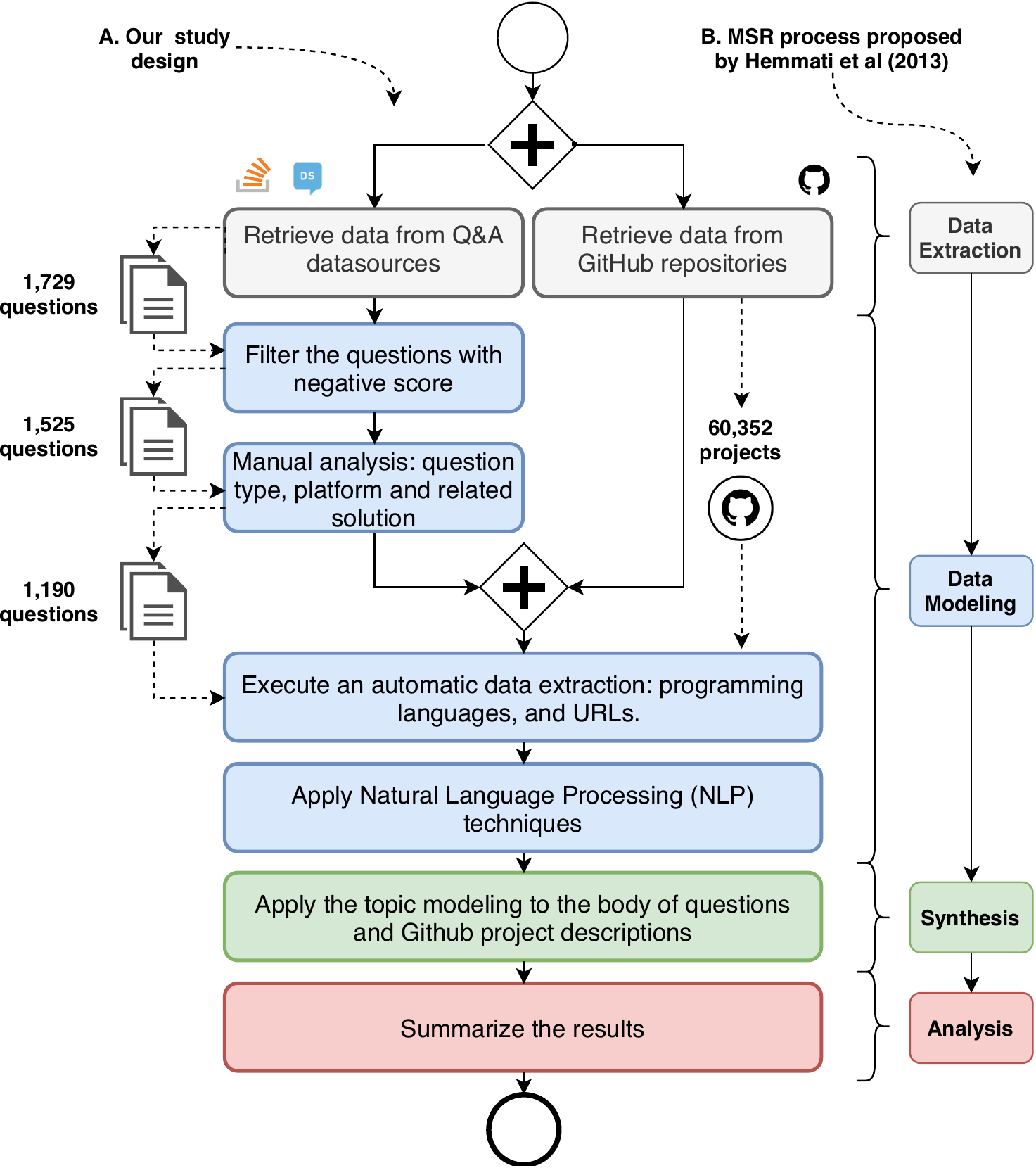}
    \caption{A. Our study design, and B. the MSR process proposed by \cite{hemmati2013msr}.}
    \label{fig_method}
\end{figure}

Figure \ref{fig_method} presents our workflow, which can be summarized in four stages: data extraction, to obtain raw data from the development artifact repositories; data modeling, related to data preparation and linking of different repositories; synthesis, to execute the topic modeling using machine learning techniques; and, analysis, to perform the interpretation of the results.

\subsection{Data extraction}
To produce meaningful information to answer our research questions, we chose relevant Q\&A websites and code repositories as data sources. The Stack Overflow represents the most important forum of discussion regarding technology and software development. We also attempted to include other databases that share the same infrastructure used by SO, but exploring other themes like the Internet of Things Q\&A, Software Engineering Q\&A, and Data Science Q\&A. However, just the Stack Overflow and Data Science Q\&A websites presented questions related to COVID-19. 

Regarding code repositories, GitHub was selected because it has one of the largest communities of developers and open source projects. Also, GitHub has a simple API to extract data, which turns it relevant to both researchers and practitioners.

To recover the questions and projects, a search string was defined considering the common terms used to characterize the COVID-19 pandemic: \textit{covid-19}, \textit{coronavirus}, \textit{2019-ncov}, \textit{sars-cov2}, and \textit{pandemic}. The questions and projects considered in this study were created from December, 2019 to May 5, 2020. Thus, it was recovered 1,729 questions and 60,352 projects.

\subsection{Data modeling}
After data extraction, it was performed four pre-processing activities. 

First, we carried out a filter to remove questions with a negative score. The Score metric (computed as \textit{up-votes} minus \textit{down-votes}) is important in Q\&A because it expresses the relevance and quality of the questions. Thus, questions with a negative score can be considered noise because the community reinforces their low quality. Other works that used Q\&A data also recommend a step to remove noise \cite{kavaler2013using, ahasanuzzaman2016mining, bandeira2019we}. This activity removed 204 questions (11.80\%).

In the second pre-processing activity, we analyzed 1,525 questions to extract the following fields: Platform, Related Solution, and Question Type. Four researchers executed this process in two steps: first, by conducting an agreement analysis in 15\% of the set; then, considering the Kappa coefficient \cite{cohen1960coefficient} of 0.636, that is considered a moderate concordance, we divided the remaining data among the researchers.

Regarding the extracted fields, we considered as \textit{Platform}: desktop, web, or mobile. In cases that were not possible to define the platform, we used the label \textit{Not Identified}. In order to set the options for \textit{Related Solution}, we conducted a pilot study to observe the most recurrent solutions discussed in our data. Thus, we used ten possibilities for this field: data visualization, web scraping, data processing, support systems to COVID-19, machine learning, data storage, natural language processing, data mining, recommender system, and others. This last one was used to the undefined \textit{Related Solution}. 

The last field is the question type and it was based on the taxonomy proposed by \cite{treude2011programmers}. This taxonomy is widely adopted in this kind of work. It has ten categories: \textbf{how-to}, to questions that ask for instructions; \textbf{discrepancy}, for questions that present unexpected results; \textbf{environment}, for questions related to development environments; \textbf{error}, to questions that include a specific error message; \textbf{decision help}, for questions that seek opinions about aspects of the development; \textbf{conceptual}, for abstract questions related to the concepts of the area; \textbf{review}, for questions that request the review of code snippets; \textbf{non-functional}, for questions about non-functional requirements; \textbf{novice}, for questions clearly asked by novice developers; and, \textbf{noise}, for questions not related to COVID-19.

At the end of the second activity, the number of questions selected for this study was 1,190. We removed questions classified as noise and questions deleted for moderation or voluntarily removed by its author.

The third activity of this stage was executing an automatic data extraction to obtain the programming languages. This identification was carried out using a term dictionary built from the association of technologies with the most used programming languages. 

Finally, in the last activity, we used Natural Language Processing (NLP) techniques to reduce noise and improve the results obtained by the Latent Dirichlet Allocation (LDA) algorithm \cite{blei2003latent}. These techniques include removing code snippets, non-ASCII characters, punctuation, words with less than three characters, and non-English texts. For stopword removal, we used a set of 1,043 terms adapted from the list proposed by \cite{puurula-2013-cumulative}. We also used WordNetLemmatizer from Python NLTK\footnote{Python NLTK website: \href{https://www.nltk.org}{https://www.nltk.org}.} to remove inflected forms of a word and performed a pruning of words with a frequency less than 5\% and more than 80\%.

\subsection{Synthesis}
This stage is responsible for the execution of methods or techniques in order to extract patterns that can be used to answer the research questions. In this case, we decided to apply the LDA algorithm in the textual data and then build a set of data visualizations to improve the data analysis.

Regarding LDA, it can identify word sets (topics), semantically coherent, to represent a textual data set and, hence, get insights about it \cite{mei2008topic}. Thus, it was used a Python version of this algorithm shared by the Scikit-learn toolkit \cite{scikit-learn}. The number of topics was empirically defined by the researchers after meetings to discuss the most suitable one, \textit{i.e.}, the number that provides a holistic overview of the data, minimizing doubts in the interpretation process. The number of topics for the Q\&A \textit{corpus} was seven, and it was eleven for the \textit{corpus} composed by descriptions of GitHub projects.

\subsection{Analysis}
The last stage addresses the review and interpretation of the results. All of them were analyzed by five researchers.

To analyze the raw data, visualizations built with Tableau\footnote{Tableau website: \href{https://www.tableau.com}{https://www.tableau.com}.} were used. Concerning the LDA models analysis, we decided to use the LDAvis method \cite{sievert2014ldavis}. This method provides an interactive view of LDA models to reduce the intrinsic complexity in the interpretation process. To prevent the overlapping of opinions during the topic discussion, each researcher performed the analysis independently. Then, in a meeting with all researchers, the results were consolidated. All topics were preserved. There was no need to join or split any of them.

\section{Results and Discussion}
\label{sec:results}
In this section, we present and discuss the results achieved in this work. Initially, we got 1,729 questions from two data sources: Stack Overflow and Data Science Q\&A websites. These questions were filtered to get 1,190 ones related to COVID-19. In parallel, 60,352 GitHub projects also associated with COVID-19 were recovered. This massive amount of data was processed to extract meaningful information to answer the research questions.

\begin{figure*}[h]
    \centering
    \includegraphics[width = 1.6\columnwidth]{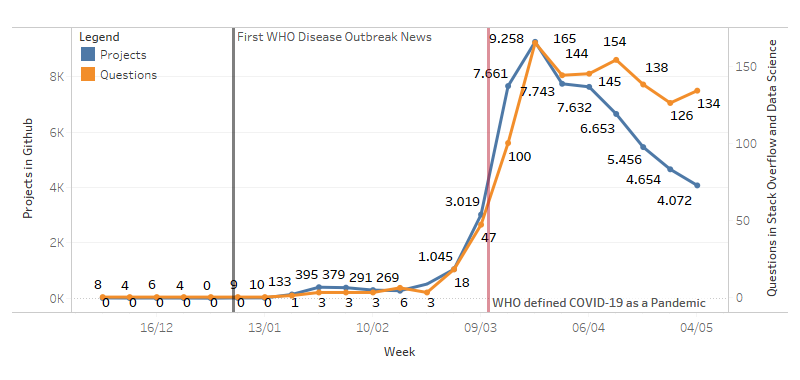}
    \caption{Distribution of Questions and Projects Related to the COVID-19.}
    \label{fig1}
\end{figure*}

\subsection{RQ1: What is the temporal distribution of questions and projects considering the COVID-19 pandemic timeline?}

This RQ is related to the temporal distribution of the questions and projects. It aims to understand the evolution of these numbers as the pandemic progressed around the world.

Figure \ref{fig1} shows the evolution of the number of questions and projects from the first week of December to the first week of May. It is perceived that the two curves (number of projects and the number of questions) have a similar behavior until the peak, which occurred on March 23rd.

We also realized that some projects started to be created even before the first WHO disease outbreak news on January 5th. Probably, because the new coronavirus was already known in China. Among these 22 projects created before January 5th, 2020, it stands out the \href{https://github.com/Perishleaf/data-visualisation-scripts}{\textit{Perishleaf/data-visualisation-scripts}} repository due to the high number of forks (68) and stars (64). A Microbiology Chinese Ph.D. created this repository to share a collection of scripts for data visualization. 

Concerning the questions, the first questions were registered only on January 20th. The peak in the number of questions and projects occurred two weeks after WHO declared the COVID-19 pandemic. Since then, the number of new projects started to reduce, while the questions fluctuated around an average of 140 new registers per week.

The behavior of both curves -- plotted in Figure \ref{fig1} -- shows a significant mobilization for developing technological solutions that could help the fight against the new coronavirus. Besides, despite the number of projects showing a downward trend, it is possible to observe that this topic will still be discussed even after the first pandemic peak.

\begin{shaded}
    RQ1 formal answer is that the number of questions and projects grew during the pandemic beginning, reaching its highest point in the week of March 11th. After that, the number of new projects has been decreasing, and the new questions are fluctuating around 140 by week.
\end{shaded}

\subsection{RQ2: What are the characteristics of the questions asked by developers in the context of the COVID-19?}

To answer this question, it was analyzed three dimensions: the question type \cite{treude2011programmers}, the platform used by the developer (desktop, Web or mobile), and the related solution discussed in the question (data visualization, web scraping, data processing, a support system to COVID-19, machine learning, data storage, natural language processing, data mining, recommender system, and others). It is also important to emphasize that most questions selected in this study come from the Stack Overflow website (1,176), but there are 14 questions from the Data Science Q\&A website.

\begin{figure*}[h]
    \centering
    \includegraphics[width = 1.4\columnwidth]{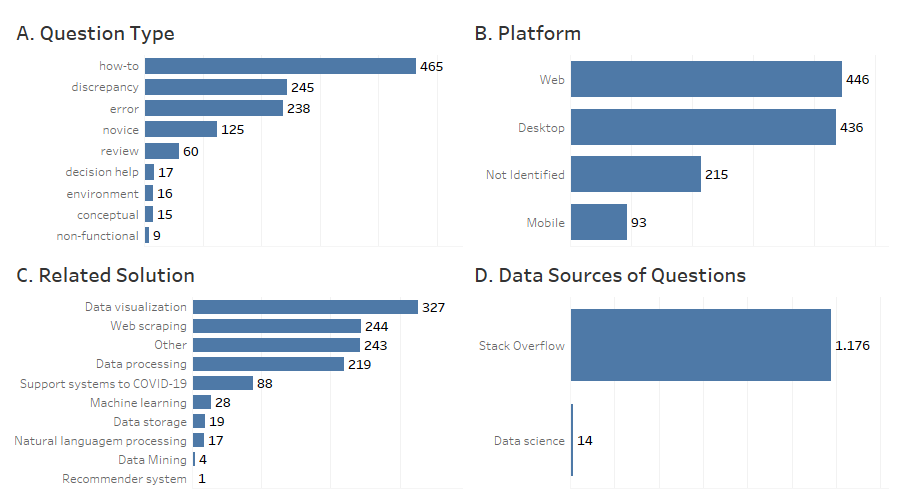}
    \caption{Number of Questions by A) Question Type, B) Platform, C)  Related Solution, and D) Data Source of Questions.}
    \label{fig2}
\end{figure*}

Figure \ref{fig2}.A presents the types of questions. Most questions were classified as the how-to type. This higher number is related to objective and direct questions about how to execute a specific task. For example, \href{http://datascience.stackexchange.com/questions/69709}{\textit{How do I test a difference between two proportions representing fatality rate for COVID 19 in Philippines and World (except Philippines)?}}.

After the how-to type, there are discrepancy, error, and novice as the main type of questions. Discrepancy and error are closely related. Questions classified as discrepancy present unexpected results, while error questions describe specific error messages. For the novice type, there are discussions that were clearly started by beginners (\textit{e.g.}, the question \href{https://stackoverflow.com/questions/60793702}{\textit{ASP.Net Core Import JSON from external API}}, in which the user starts with ``\textit{I am new to both .NET Core}''). The number of questions in these three categories (discrepancy, error, and novice) and the descriptions observed by the researchers indicate that many people with an aptitude for software development but with little experience have decided to work on voluntary projects related to COVID-19. For example, \href{https://stackoverflow.com/questions/61294586}{\textit{How can you customize the contents of popups when you click on built-in locations in a map from the Google Maps API?}}, in which the user presented itself as a COVID-related volunteer working on a solution with the Google Maps API.

In Figure \ref{fig2}.B, we can observe the number of questions by platform, and in Figure \ref{fig2}.C the distribution by the related solution. With these two graphics, we realized that most questions are inserted into the context of data visualization and web scraping on the Web platform. However, a significant number of questions related to data processing were found on the Desktop platform.

About these data, it was expected a large number of how-to questions focused on data visualization for the Web platform. This behavior was expected due to the fundamental role that useful visualizations play in planning actions against Sars-CoV-2. We also expected a high number of questions reporting difficulties faced by novice users, like errors or discrepancies. However, the low number of discussions focused on the application of Machine Learning, Data Mining, and Natural Language Processing caught attention. These techniques can be used to predict events, to support diagnosis, and to generate information from raw data, strengthening strategies to combat the pandemic. This fact is particularly strange since most of the projects found in GitHub have as their associated topic the development of solutions based on machine learning to tackle the pandemic problem (this topic is further discussed in the following subsections). We believe that SO is not a specific site for discussing machine learning questions, being more used for questions related to software development in general, especially for novice programmers.

\begin{shaded}
    We can answer the RQ2, stating that the most frequent types of questions asked are how-to, discrepancy, error, and novice. In general, these questions are related to data visualization and Web scraping solutions for the Web platform. Until the moment when the data were extracted for this research, the number of questions about Machine Learning, Data Mining, and Natural Language Processing was low.
\end{shaded}

\subsection{RQ3: Which programming languages are being used to develop solutions related to the COVID-19?}

\begin{figure}[h]
    \centering
    \includegraphics[width = 0.9\columnwidth]{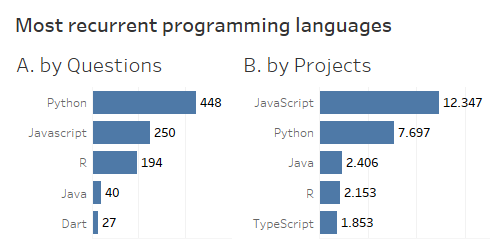}
    \caption{Most recurrent programming languages: A) by questions and B) by Projects.}
    \label{fig4}
\end{figure}

Regarding programming languages, Figure \ref{fig4} shows the most recurrent languages: A) by questions and B) by projects. In this way, a correlation was observed between the most discussed languages and the projects created on GitHub.

The four most discussed and used languages were Python, Javascript, Java, and R. These languages are directly related to the solutions that are being developed during this pandemic. For example, the R language is mostly used for data analysis and processing; Python has been widely used for creating web scraping and for data analysis on Jupyter notebooks\footnote{Jupyter website: \href{https://jupyter.org/}{https://jupyter.org/}.}; Javascript is commonly used for developing Web solutions as data visualizations with D3.js\footnote{D3.js website: \href{https://d3js.org/}{https://d3js.org/}.}; and Java can be used for both Web and mobile development.

The choice of the programming language is also linked with the maturity of the tools used for a specific task. Figure \ref{fig5} presents the most discussed technologies considering the tags assigned for the questions. In this ranking, it is possible to observe tools, frameworks, and libraries written in Python such as: pandas\footnote{Pandas website: \href{https://pandas.pydata.org/}{https://pandas.pydata.org/}.}, for data manipulation, beautifulsoup\footnote{Beautiful Soup library: \href{https://pypi.org/project/beautifulsoup4/}{https://pypi.org/project/beautifulsoup4/}.} to scrape information from web pages, and Matplotlib\footnote{Matplotlib: \href{https://matplotlib.org/}{https://matplotlib.org/}.} for creating static, animated, and interactive visualizations. Related to the use of Javascript (\textit{e.g.}, the main used packages are: React.js\footnote{React.js: \href{https://reactjs.org/}{https://reactjs.org/}.}, a library to build user interfaces; Node.js\footnote{Node.js: \href{https://nodejs.org/}{https://nodejs.org/}.}, a platform to support the development of server-side JS apps; and plotly\footnote{Plotly webpage: \href{https://plotly.com/javascript/}{https://plotly.com/javascript/}.}, a charting library built on top of D3.js. The technology related to Java is selenium\footnote{Selenium: \href{https://www.selenium.dev/}{https://www.selenium.dev/}.}, used to create capture and replay tasks in browsers; in R there is the use of ggplot2\footnote{ggplot2 website: \href{https://ggplot2.tidyverse.org/}{https://ggplot2.tidyverse.org/}.}, a visualization package and the use of shiny\footnote{Shiny package: \href{https://shiny.rstudio.com/}{https://shiny.rstudio.com/}.}, for interactive web apps.

\begin{figure}[h]
    \centering
    \includegraphics[width = 0.7\columnwidth]{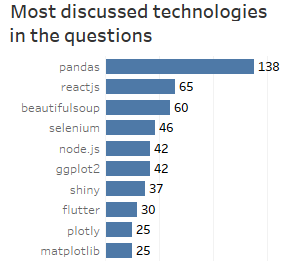}
    \caption{The most discussed technologies.}
    \label{fig5}
\end{figure}

\begin{shaded}
    For the RQ3, we can state that the most used and discussed programming languages are Python, Javascript, Java, and R. This reflects the great interest in solutions focused on data visualization, web scraping, and data processing. The choice for these languages is also directly linked to the maturity of specific libraries and tools created for each of them.
\end{shaded}

\subsection{RQ4: What are the hot topics considering the questions and project descriptions in the context of the COVID-19 pandemic?}

Finding the most relevant topics in a specific area can assist in identifying trends and opportunities. In this work, it was applied a topic modeling algorithm, called LDA, in two data sources: i) the body of the questions from Stack Overflow and Data Science websites, and ii) the description of the projects in GitHub. The execution of this algorithm required a list of NLP pre-processing tasks (described in Subsection II.B). As a result, seven (7) hot topics were obtained for the questions, and eleven (11) for the projects.

Figure \ref{fig5} shows a static view of the topics for each data source after the interpretation process. However, it is possible to get an interactive view of these topics accessing the link \href{https://lostufpi.github.io/covid19-se/datavis}{https://lostufpi.github.io/covid19-se/datavis}.

\begin{figure}[h]
    \centering
    \includegraphics[width = \columnwidth]{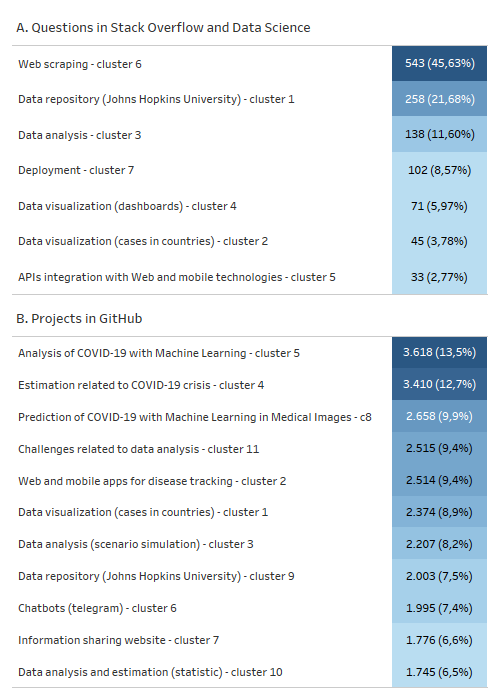}
    \caption{Most discussed topics considering A) the questions in Stack Overflow and Data Science; and B) the projects in GitHub.}
    \label{fig5}
\end{figure}

Regarding the questions, the three most discussed topics are about \textbf{web scraping} (with 45.63\% of the questions were classified in this topic), \textbf{data repository of Johns Hopkins University} (with 21.68\%), and \textbf{data analysis} (11.60\%).

Concerning the projects, the three most discussed topics are \textbf{analysis of COVID-19 with Machine Learning} (with 3,618 projects), \textbf{estimation of COVID-19 crisis/impact} (with 3,410 projects), and \textbf{prediction of COVID-19 with Machine Learning in Medical Images} (with 2,658 projects). It is important to note that it was necessary to remove 33,537 projects (55.57\%) with blank descriptions or that were not written in English for this analysis. Figure \ref{fig5}.B presents the results achieved, considering 26,815 projects (44.43\% of the total).

These results present an interesting divergence. Considering the questions analyzed, the most relevant topics focus on data acquisition and analysis. However, the most prominent topics obtained from the projects' description indicate a large number of initiatives that try to use Machine Learning to analyze  COVID-19 data. A hypothesis that could justify this divergence would be the fact that Q\&A websites, especially SO, are mostly used by developers with less experience. This profile is usually not associated with more advanced programming subjects, like Machine Learning and Data mining.

\begin{shaded}
    RQ4 answer is that regarding questions, the most recurrent topics are: web scraping, data repository (Johns Hopkins), and data analysis and prediction. This indicates that a significant number of questions are focused on COVID-19 data acquisition, analysis, and prediction; and reinforces the relevance of the data repository maintained by Johns Hopkins University. In contrast, the hot topics modeled from GitHub projects' description are analysis of COVID-19 with Machine Learning, estimation of related COVID-19 crisis, and prediction of COVID-19 with Machine Learning in medical images. This divergence suggests that many initiatives seek to develop smart solutions to tackle the pandemic and that these solutions are not often discussed on websites like Stack Overflow.
\end{shaded}

\subsection{RQ5: Which GitHub projects related to the COVID-19 context stand out?}

This RQ guided the investigation to understand the characteristics of the GitHub projects related to COVID-19. For this, it was collected the number of forks, pull requests, disk usage, commits, and collaborators of these projects.

\begin{figure}[h]
    \centering
    \includegraphics[width = \columnwidth]{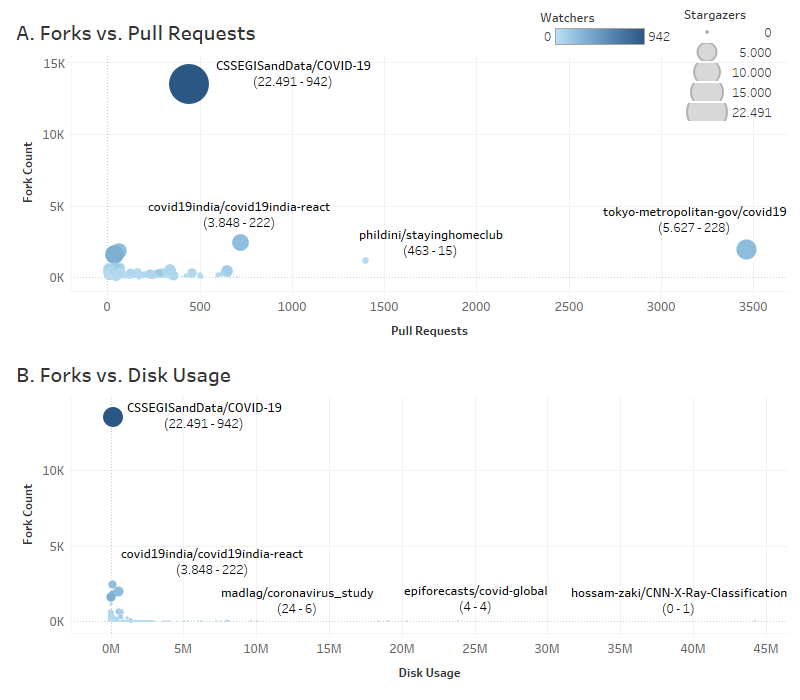}
    \caption{Scatter plot correlating A) number of forks and pull requests and B) number of forks and disk usage in the GitHub project related to COVID-19. In these visualizations, we also encoded the number of watchers in the blue color scale and the number of users who marked the project with a star in the circles' size.}
    \label{fig6}
\end{figure}

Figure \ref{fig6} presents a scatter plot correlating A. the number of forks with pull requests and B. forks with disk usage. In these charts, it is possible to observe the number of watchers, encoded in the blue color scale, and the number of users who mark the project with a star, encoded in the circles' size. 

With the visualization provided by Figure \ref{fig6}, it was possible to realize that most projects have a lower number of forks, pull requests, and disk usage. This observation indicates that there are a large number of small projects. However, some projects have a higher number of forks (\textit{i.e.}, a repository copy). For example, the COVID-19 Data Repository by the Center for Systems Science and Engineering at Johns Hopkins University (\href{https://github.com/CSSEGISandData/COVID-19}{CSSEGISandData/COVID-19}). In contrast, the repository of the Tokyo COVID-19 Task Force website (\href{https://github.com/tokyo-metropolitan-gov/covid19}{tokyo-metropolitan-gov/covid19}) has few forks, but a high number of pull requests (\textit{i.e.}, a method of submitting contributions to the project).

Regarding the correlation between forks and disk usage, we can observe that medical image processing projects require more space in the disk, even compared to data repositories. The biggest repository founded was a project to build a Convolutional Neural Network (CNN) to classify X-Ray images from both the National Institutes of Health (NIH) dataset, as well as a Kaggle COVID-19 dataset (\href{https://github.com/hossam-zaki/CNN-X-Ray-Classification}{hossam-zaki/CNN-X-Ray-Classification}).

\begin{figure}[b]
    \centering
    \includegraphics[width = \columnwidth]{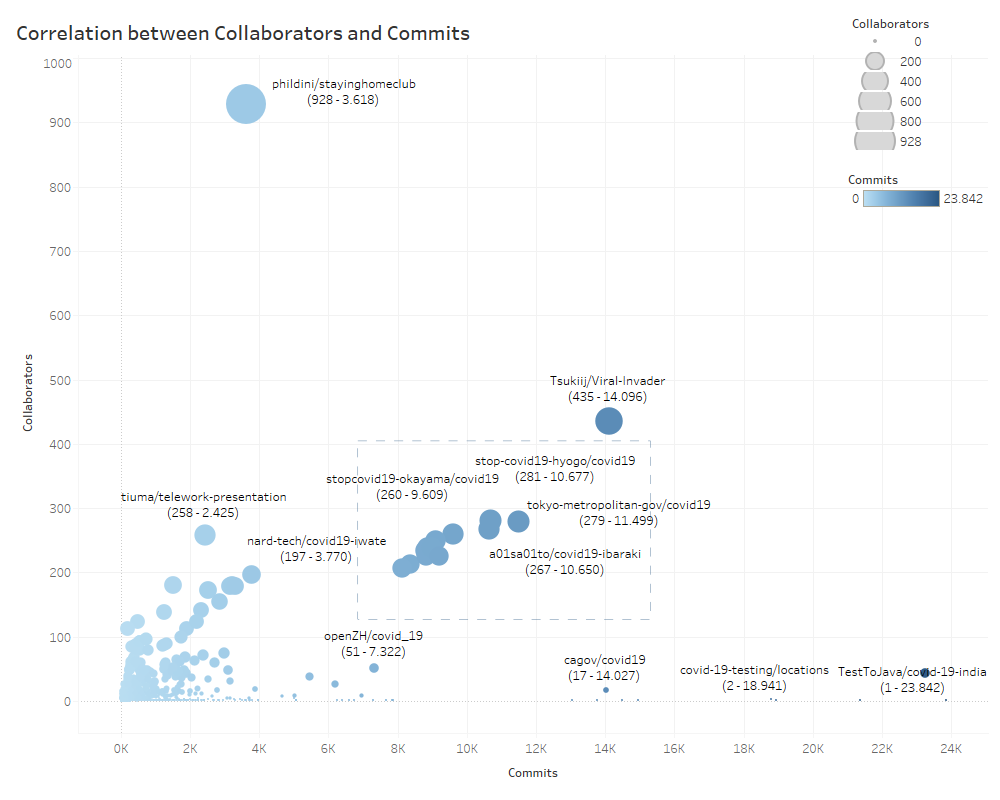}
    \caption{Correlation between collaborators and commits.}
    \label{fig8}
\end{figure}

The number of commits and collaborators was collected to analyze the team's size and the activity level of the projects. The average for collaborators is $1.384$ with a standard deviation of $6.462$, and for commits, the average is $36.28$ with $319.60$ of standard deviation. This indicates that, in general, the teams of these projects are small and perform a few commits. 59,159 projects (98\%) have teams with less than five members. Most projects (74\%) have only two members (the owner and a collaborator). 82.63\% of the projects have a number of commits less than 30. This suggests that most COVID-19 projects can be characterized by small projects with few collaborators. However, there are outliers. Twenty-eight projects have more than 100 collaborators, and their average commits are 5,767. The three projects with the largest teams are: \href{https://github.com/phildini/stayinghomeclub}{phildini/stayinghomeclub} (that provides a list of companies or events changed because of COVID-19), \href{https://github.com/Tsukiij/Viral-Invader}{Tsukiij/Viral-Invader} (with a remake of the arcade game space invaders with a COVID-19 theme), and \href{https://github.com/stop-covid19-hyogo/covid19}{stop-covid19-hyogo/covid19} (that contains the Hyogo-Japan COVID-19 website). We also analyzed the measures of commits and collaborators within each topic. However, it was not observed any significant changes.

Figure \ref{fig8} presents a scatter plot with the correlation between the number of collaborators and the number of commits. In addition to the projects already mentioned, it is important to highlight i) the \href{https://github.com/TestToJava/covid-19-india}{TestToJava/covid-19-india} project (that brings the current situation of COVID-19 in India), which has only one collaborator, but a high number of commits; and ii) Japanese task force repositories for monitoring the disease in several locations (Tokyo, Ibaraki, Hyogo, Okayama, Kagawa, Saitama, Tochigi, Kyoto) - highlighted with a dashed rectangle. These last repositories have a high number of collaborators and commits.

\begin{shaded}
    RQ5 answer is that different projects stand out in each of the measures investigated, such as the CSSEGISandData/COVID-19 project with a high number of forks, the tokyo-metropolitan-gov/covid19 project with a high number of pull requests, and the hossam-zaki/CNN-X-Ray-Classification project with high disk usage. Regarding commits and collaborators, it was found that most projects have small teams with a low number of commits. Finally, it was possible to verify that the most mentioned repository in the questions is the COVID-19 data repository of Johns Hopkins University.
\end{shaded}

\section{Validity Threats}
\label{sec:threats}
The first threat in this study is related to the selection of data sources. As data mining studies reflect the used datasets, it is essential to choose them considering the relevance to the investigated problem. Therefore, four Q\&A websites (Stack Overflow, Data Science, Internet of Things, and Software Engineering) were analyzed due to their high representativeness for software development. However, only two of them have questions related to COVID-19: Stack Overflow and Data Science. We argue that these two data sources are significant to answer our research questions because they have a large and active community, and especially the Stack Overflow data are also used in many previous scientific papers \cite{chen2019modeling, beyer2019kind, ragkhitwetsagul2019toxic, wu2019developers}.

Concerning the manual data extraction threat, we sought to reduce the bias involving four researchers to conduct this process. At the beginning of the analysis, two preliminary rounds were conducted to assess the evaluators' agreement. As a result, we obtained a Kappa coefficient of 0.636. It was then decided to proceed with the extraction because the disagreements will be resolved in meetings with all researchers.

Another threat arose during the topic modeling. This activity has threats due to the empirical nature of the definition of the number of topics and topic interpretation. Thus, the number of topics were chosen considering the trade-off between the model complexity and its representativeness. The interpretations were defined in a meeting with the four researchers that performed the manual analysis of the questions and one more expert researcher. At this point, the LDAvis method was used to support this activity because it allows an interactive view of the multidimensional model generated by the LDA.

\section{Conclusions}
\label{sec:conclusion}
The world is going through something new for the vast majority of people. An illness completely changed our way of life, imposing severe crowding restrictions and close contact with loved ones. Several essential activities, such as education, commerce, and tourism, have stopped for a few months and still remain in many countries.

During that time, many people tried to help fight this disease using their weapons. People connected to computing started to develop solutions that somehow help to tackle the pandemic. This study investigated exactly what has been done by people connected with computing. For this, data mining was performed in some repositories, which have considerable relevance to the computing area: the Stack Overflow and Data Science Q\&A websites, as well as the GitHub code repository.

We analyzed 1,190 questions from Stack Overflow and Data Science Q\&A and 60,352 GitHub projects. We have found 18 meaningful topics (seven from questions and eleven from project descriptions) about several themes. We also realized that questions and projects grew throughout the pandemic and that there is a correlation between them. The main questions related to the \-coronavirus were classified as how-to, concerning web scraping and data visualization/processing, using Python, JavaScript, and R. The repositories in GitHub are in the majority of Machine Learning projects, using JavaScript, Python, and Java.

It is noticeable people's interest in helping to tackle the disease in whatever way possible. The study results showed that many beginners and volunteers are working hard to try to build solutions that can contribute to this cause. Also, the identified trends can be used to reinforce quick response actions in future pandemics. However, we are still far from a full victory against this virus. Therefore, the data collected in this work can support the development of solutions by researchers and practitioners engaged in combating COVID-19, indicating the most used data repositories, recurring technologies, and programming languages, open-source projects that are standing out, in addition to the most discussed subjects. An interactive data visualization page was created and made available for public access as an additional contribution to the work. For future works, we will analyze the synergism between software communities and mechanisms to improve their interaction in global mobilization cases.

\section*{Code and Data Availability} 
In order to improve the work reproducibility, all codes and data used are publicly available on the Internet. The query to search the original questions can be accessed at the link: \href{https://data.stackexchange.com/stackoverflow/query/1232694/my-coronavirus-query}{data.stackexchange.com/stackoverflow/query/1232694}. The codes used to build the LDA models with the Scikit-learn, and the visualizations presented in this paper are available on GitHub repository: \href{https://github.com/lostufpi/covid19-se}{https://github.com/lostufpi/covid19-se}.

\section*{Acknowledgments}
The authors would like to thank CNPQ for the Productivity Scholarship of Pedro A. dos Santos Neto DT-2 ($N^{o}$ 315198 / 2018-4) and for the Productivity Scholarship of Rossana M. C. Andrade DT-2 ($N^{o}$ 315543 / 2018-3).



\begin{thebibliography}{10}
\providecommand{\url}[1]{#1}
\csname url@samestyle\endcsname
\providecommand{\newblock}{\relax}
\providecommand{\bibinfo}[2]{#2}
\providecommand{\BIBentrySTDinterwordspacing}{\spaceskip=0pt\relax}
\providecommand{\BIBentryALTinterwordstretchfactor}{4}
\providecommand{\BIBentryALTinterwordspacing}{\spaceskip=\fontdimen2\font plus
\BIBentryALTinterwordstretchfactor\fontdimen3\font minus
  \fontdimen4\font\relax}
\providecommand{\BIBforeignlanguage}[2]{{%
\expandafter\ifx\csname l@#1\endcsname\relax
\typeout{** WARNING: IEEEtran.bst: No hyphenation pattern has been}%
\typeout{** loaded for the language `#1'. Using the pattern for}%
\typeout{** the default language instead.}%
\else
\language=\csname l@#1\endcsname
\fi
#2}}
\providecommand{\BIBdecl}{\relax}
\BIBdecl

\bibitem{munster2020novel}
V.~J. Munster, M.~Koopmans, N.~van Doremalen, D.~van Riel, and E.~de~Wit, ``A
  novel coronavirus emerging in china—key questions for impact assessment,''
  \emph{New England Journal of Medicine}, vol. 382, no.~8, pp. 692--694, 2020.

\bibitem{hemmati2013msr}
H.~Hemmati, S.~Nadi, O.~Baysal, O.~Kononenko, W.~Wang, R.~Holmes, and M.~W.
  Godfrey, ``The msr cookbook: Mining a decade of research,'' in \emph{2013
  10th Working Conference on Mining Software Repositories (MSR)}.\hskip 1em
  plus 0.5em minus 0.4em\relax IEEE, 2013, pp. 343--352.

\bibitem{kitchenham2015evidence}
B.~A. Kitchenham, D.~Budgen, and P.~Brereton, \emph{Evidence-based software
  engineering and systematic reviews}.\hskip 1em plus 0.5em minus 0.4em\relax
  CRC press, 2015, vol.~4.

\bibitem{kavaler2013using}
D.~Kavaler, D.~Posnett, C.~Gibler, H.~Chen, P.~Devanbu, and V.~Filkov, ``Using
  and asking: Apis used in the android market and asked about in
  stackoverflow,'' in \emph{International Conference on Social
  Informatics}.\hskip 1em plus 0.5em minus 0.4em\relax Springer, 2013, pp.
  405--418.

\bibitem{ahasanuzzaman2016mining}
M.~Ahasanuzzaman, M.~Asaduzzaman, C.~K. Roy, and K.~A. Schneider, ``Mining
  duplicate questions of stack overflow,'' in \emph{2016 IEEE/ACM 13th Working
  Conference on Mining Software Repositories (MSR)}.\hskip 1em plus 0.5em minus
  0.4em\relax IEEE, 2016, pp. 402--412.

\bibitem{bandeira2019we}
A.~Bandeira, C.~A. Medeiros, M.~Paixao, and P.~H. Maia, ``We need to talk about
  microservices: an analysis from the discussions on stackoverflow,'' in
  \emph{Proceedings of the 16th International Conference on Mining Software
  Repositories}.\hskip 1em plus 0.5em minus 0.4em\relax IEEE Press, 2019, pp.
  255--259.

\bibitem{cohen1960coefficient}
J.~Cohen, ``A coefficient of agreement for nominal scales,'' \emph{Educational
  and psychological measurement}, vol.~20, no.~1, pp. 37--46, 1960.

\bibitem{treude2011programmers}
C.~Treude, O.~Barzilay, and M.-A. Storey, ``How do programmers ask and answer
  questions on the web?: Nier track,'' in \emph{2011 33rd International
  Conference on Software Engineering (ICSE)}.\hskip 1em plus 0.5em minus
  0.4em\relax IEEE, 2011, pp. 804--807.

\bibitem{blei2003latent}
D.~M. Blei, A.~Y. Ng, and M.~I. Jordan, ``Latent dirichlet allocation,''
  \emph{Journal of machine Learning research}, vol.~3, no. Jan, pp. 993--1022,
  2003.

\bibitem{puurula-2013-cumulative}
\BIBentryALTinterwordspacing
A.~Puurula, ``Cumulative progress in language models for information
  retrieval,'' in \emph{Proceedings of the Australasian Language Technology
  Association Workshop 2013 ({ALTA} 2013)}, Brisbane, Australia, Dec. 2013, pp.
  96--100. [Online]. Available: \url{https://www.aclweb.org/anthology/U13-1013}
\BIBentrySTDinterwordspacing

\bibitem{mei2008topic}
Q.~Mei, D.~Cai, D.~Zhang, and C.~Zhai, ``Topic modeling with network
  regularization,'' in \emph{Proceedings of the 17th international conference
  on World Wide Web}, 2008, pp. 101--110.

\bibitem{scikit-learn}
F.~Pedregosa, G.~Varoquaux, A.~Gramfort, V.~Michel, B.~Thirion, O.~Grisel,
  M.~Blondel, P.~Prettenhofer, R.~Weiss, V.~Dubourg, J.~Vanderplas, A.~Passos,
  D.~Cournapeau, M.~Brucher, M.~Perrot, and E.~Duchesnay, ``Scikit-learn:
  Machine learning in {P}ython,'' \emph{Journal of Machine Learning Research},
  vol.~12, pp. 2825--2830, 2011.

\bibitem{sievert2014ldavis}
C.~Sievert and K.~Shirley, ``Ldavis: A method for visualizing and interpreting
  topics,'' in \emph{Proceedings of the workshop on interactive language
  learning, visualization, and interfaces}, 2014, pp. 63--70.

\bibitem{chen2019modeling}
H.~Chen, J.~Coogle, and K.~Damevski, ``Modeling stack overflow tags and topics
  as a hierarchy of concepts,'' \emph{Journal of Systems and Software}, vol.
  156, pp. 283--299, 2019.

\bibitem{beyer2019kind}
S.~Beyer, C.~Macho, M.~Di~Penta, and M.~Pinzger, ``What kind of questions do
  developers ask on stack overflow? a comparison of automated approaches to
  classify posts into question categories,'' \emph{Empirical Software
  Engineering}, pp. 1--44, 2019.

\bibitem{ragkhitwetsagul2019toxic}
C.~Ragkhitwetsagul, J.~Krinke, M.~Paixao, G.~Bianco, and R.~Oliveto, ``Toxic
  code snippets on stack overflow,'' \emph{IEEE Transactions on Software
  Engineering}, 2019.

\bibitem{wu2019developers}
Y.~Wu, S.~Wang, C.-P. Bezemer, and K.~Inoue, ``How do developers utilize source
  code from stack overflow?'' \emph{Empirical Software Engineering}, vol.~24,
  no.~2, pp. 637--673, 2019.

\end{thebibliography}
\end{document}